\documentclass[noeprint,twocolumn,showpacs,amsmath,amssymb,sort&compress,floatfix,superscriptaddress,prl,aps]{revtex4-2}

\usepackage{graphicx}
\usepackage{dcolumn}
\usepackage[normalem]{ulem}
\usepackage{bm}
\usepackage[hypertexnames=false]{hyperref}
\usepackage[mathlines]{lineno}
\usepackage{mathtools}
\usepackage{amsmath}
\usepackage{graphicx}

\usepackage[caption=false]{subfig}
\usepackage[english]{babel}
\usepackage{xcolor}
\usepackage{bm}
\usepackage{xparse}
\addto\captionsenglish{}
\usepackage{flushend}
\usepackage{nowidow}
\usepackage{balance}
\usepackage{placeins}
\usepackage{verbatim}
\usepackage{tabularx}

\newcommand{\fig}{Fig.~}
\newcommand{\figs}{Figs.~}
\newcommand{\eqn}{Eq.~}

\newcommand{\pd}{\partial}
\newcommand{\abs}[1]{\left|#1\right|}
\newcommand{\avg}[1]{\left\langle#1\right\rangle}
\renewcommand{\vec}[1]{\bm{#1}}
\newsavebox{\slantbox}
\newcommand{\msymv}{\text{\scalebox{0.7}{\slshape V}}}

\newcommand{\ueph}{SUPA, School of Physics and Astronomy, University of Edinburgh, Peter Guthrie Tait Road, Edinburgh EH9 3FD, United Kingdom}

\makeatletter
\def\maketitle{
\@author@finish
\title@column\titleblock@produce
\suppressfloats[t]}
\makeatother

\newcommand{\papertitle}
{Flow of deformable droplets: self-pinned glasses and string-like flow}

\begin{document}
\title{\papertitle}

\author{Achille Quarante}
\affiliation{\ueph}
\author{Michael Chiang}
\affiliation{\ueph}
\author{Davide Marenduzzo}
\affiliation{\ueph}
\author{Giuseppe Negro}
\affiliation{\ueph}

\begin{abstract}
We investigate, through numerical simulations, the rheology of a dry suspension of deformable droplets under pressure-driven flow. The system exhibits two force-driven dynamical transitions. At low forcing, the suspension behaves as a yield-stress material: below a critical force, droplets remain arrested in an amorphous solid-like state. Our simulations suggest that yielding is controlled by droplet contacts and predict that the critical force strongly depends on deformability. Above yielding, the suspension does not flow steadily but rather enters an intermittent, stick-slip regime characterised by long-lived caging and non-Gaussian velocity fluctuations. This state can be interpreted as a ``self-pinned'' glass, in which slowly evolving droplet overlaps generate an effective rugged energy landscape that dynamically traps droplets and produces intermittent rearrangements reminiscent of near-critical dynamics in depinning models. At larger forcing, droplets deform sufficiently to continuously exchange neighbours, progressively annealing the overlap structure and driving a dynamic transition to a string-like, flowing state. Our results identify the restructuring of overlap networks as a generic mechanism which controls flow in driven suspensions of deformable particles.
\end{abstract}
\maketitle

\textit{Introduction}---How complex fluids respond to external driving is a central question in soft matter physics, with implications ranging from industrial processing to biological function~\cite{Larson1999,Schaefer2022,Kayal2026}. In these fluids, macroscopic rheology is governed not by molecular interactions alone but by mesoscopic structure, the collective organisation of droplets, bubbles, or cells, whose morphology evolves under flow~\cite{Larson1999,Mason1995}. An important manifestation of this flow--structure coupling is the yield stress~\cite{Bonn2017,Balmforth2014}: a critical stress threshold that must be overcome for the material to flow, exhibited by colloidal suspensions~\cite{Bonn2017,Aime2023,He2025,Koumakis2013}, dense emulsions~\cite{Mason1996,Knowlton2014,Robert2018}, foams~\cite{Heller1987,Princen1983,Khan1986,Pratt2003,Katgert2009,Kraynik1998,Schott2025}, and gels~\cite{Coussot2014,Patrick2025,Laurati2009}. Here, yielding is not a property of the individual constituents but emerges from the collective evolution of mesoscopic contacts, deformations, and rearrangements under external forcing~\cite{Cates1998,Benzi2014,Benzi2016,Lulli2018,Fielding2023,Scherrer2025}. 

Colloidal suspensions have been widely studied as model systems for complex fluid rheology, showing a rich phenomenology that includes shear thinning, shear thickening, and glass-like arrest at high packing fractions~\cite{Pusey1986, Jones1991, vanderWerff1989,Sollich1997, Cheng2011}. These behaviours have been rationalised within the framework of mode-coupling theory and kinetic elasto-plastic models~\cite{Bouchaud1996,Fuchs2009,Nicolas2018}, and scaling theories~\cite{Lin2014}. Yet, hard-sphere colloids are essentially rigid, and the role of particle \textit{deformability}---central to emulsions~\cite{Zinchenko_Davis_2017}, foams~\cite{Heller1987,Princen1983,Khan1986,Pratt2003,Katgert2009,Kraynik1998,Manning2023}, and biological cell suspensions~\cite{Michelot2022,Bashir2022,Manning2022,ladoux2015,nejad2024,vandenBerg2025,BiDepeng2016,Giomi20223,Manning2023}---remains comparatively less explored,  and a unified picture of the mechanics and rheological response of both rigid and deformable particles under shear is still lacking. 

Recently, numerical studies of deformable droplet suspensions have begun to bridge this gap~\cite{Foglino2017,Foglino2018,Fei2020,Negro2023,hadjifrangiskou2026shear}. These works showed that deformability introduces new phenomena absent in hard-sphere systems: discontinuous shear thinning associated with a nonequilibrium transition between a hard and a soft, shape-fluctuating phase~\cite{Foglino2017, Foglino2018}, and a yield-stress transition linked to droplet-droplet overlaps, which requires strict area conservation to exist~\cite{Negro2023}. However, the full picture of how deformability controls yielding and the mechanics of post-yielding morphology remains an open question.

Here, we show that upon increasing forcing, a dry deformable suspension undergoes two dynamical transitions. First, the system yields from a solid phase to a stick-slip regime and behaves like a ``self-pinned'' glass. This regime arises as droplet overlaps persist for a long time and are dynamically quenched, acting as a random potential that traps the droplets in long-lived local cages before they elastically snap to the next cage. This behaviour is analogous to that observed in the Prandt-Tomlinson model~\cite{Prandtl1928} near the depinning transition, where local energy barriers in an external periodic potential lead to cage-and-slip dynamics associated with barrier hopping. In our case, the emerging potential is self-created by the overlaps formed dynamically and thus is not external. Upon larger forcing, the droplets deform sufficiently to exchange neighbours and flow in a string-like fashion without being slowed down by overlaps, losing the intermittent dynamics. 

\textit{Model}---We study the rheology of a suspension of $N$ soft droplets under an applied flow in a square channel using a multiphase-field approach~\cite{Foglino2017,Negro2023} [\fig\ref{fig:model}(a), left]. Each droplet has a radius $R$ and is modelled by a phase field $\phi_i$ ($i = 1, ..., N$). The morphology of individual droplets is governed by the following free energy:
\begin{equation}
\begin{aligned}
    \mathcal{F} = \sum_{i=1}^N \Bigg[
    & \int d^2\vec{r}\,\left[\frac{\alpha}{4}\phi_i^2(\phi_i-\phi_0)^2
    + \frac{\kappa}{2}(\vec{\nabla} \phi_i)^2\right] \\
    & + \epsilon \sum_{i<j=1}^N \int d^2\vec{r}\,\phi_i^2\phi_j^2
    \Bigg] \;.
\end{aligned}
\label{eqn:free_energy}
\end{equation}
Here, the first term helps define the droplet shape, where $\phi_i$ = $\phi_0 = 2$ represents its interior and $\phi_i = 0$ its exterior. The second term regulates the droplet interface and favours circular droplets by minimising the interfacial tension $\sigma = \sqrt{(8\kappa \alpha)/9}$, while each droplet's interface is defined by $\xi = \sqrt{2\kappa/\alpha}$~\cite{Pagonabarraga2002}. The final term imposes steric repulsion between droplets (strength $\epsilon$) by penalising droplet overlaps. The parameters above allow us to define a dimensionless quantity $d = \epsilon/\alpha$ that controls the droplet's deformability: for $d \ll 1$ droplets tend to be circular and overlap with their neighbours, whereas for $d \gg 1$ they tend to deform and minimise overlaps~\cite{Loewe2020}.

To obtain dynamics, the phase fields $\phi_i$ are evolved through a set of Cahn-Hilliard equations
\begin{equation}
    \frac{\partial \phi_i}{\partial t} + \vec{v} \cdot \vec{\nabla} \phi_i = M \nabla^2 \frac{\delta \mathcal{F}}{\delta \phi_i} \;,
    \label{eqn:cahn-hillard}
\end{equation}
where $M$ is the mobility, $\vec{v} = v_x(y)\,\hat{\vec{e}}_x$ the flow velocity, and $\delta \mathcal{F}/\delta \phi_i$ the chemical potential of the $i$th droplet.

To focus on how interactions between droplets and their geometry influence the yielding behaviour, we simulate the droplets in a ``dry'' context by imposing a steady-state Poiseuille flow and neglecting hydrodynamic effects [\fig\ref{fig:model}(a), right]. The fluid velocity profile thus reads 
\begin{equation}
    \label{eqn:poiseuile_flow}
    v_x(y) = \frac{4v_{\text{max}}}{ L^2}y(L-y) \;,
\end{equation}
where $v_{\text{max}}$ is the maximum flow speed at the centre of the channel of width $L$. This allows us to define a dimensionless P{\'e}clet number $\text{Pe} = v_{\text{max}}/V$ that we vary in the simulations, where $V = M\epsilon/L$ is a typical velocity of droplets arising from passive thermodynamic forces.  

We impose neutral wetting boundary conditions for the droplets (i.e., $\pd_y\nabla^2\phi = 0$)~\cite{Foglino2017,Foglino2018,Negro2023}, and they are initially placed randomly within the channel and are allowed to equilibrate before applying the shear flow. We solve the phase-field equations using finite-difference methods~\cite{Loewe2020,Hopkins2022}, with the full simulation procedure and parameter values used given in the Supplemental Material~\cite{SI}. We consider a packing fraction $\Phi = 0.67$ with different system sizes ($L = 128$ to $512$), all giving qualitatively similar results.

\begin{figure}[t]
    \centering
    \includegraphics[width=\columnwidth]{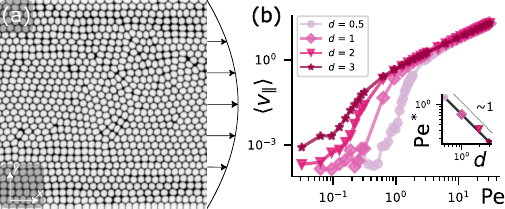}
    \caption{Yielding of a suspension of soft droplets under flow in a channel. (a) Left: A representative snapshot of the system with flow in the $x$ direction, and the parameters used are $d = 0.5$, $\text{Pe} = 23$, and $L = 512$. Right: The velocity profile of the flow. (b) Mean droplet centre-of-mass velocity $\msymv_\|$ (normalised by $V$, see~\cite{SI}) versus $\text{Pe}$ for different $d$ with $L = 128$. Inset: The critical P{\'e}clet ($\text{Pe}^*$) for yielding as a function of $d$.}
    \label{fig:model}
\end{figure}

\textit{Results}---We begin by characterising the onset of motion as the deformable suspension is subjected to increasing forcing. As P{\'e}clet rises, the system exhibits a well-defined yielding threshold $\text{Pe}^*$ below which droplets remain essentially arrested, as confirmed by both vanishing centre-of-mass velocity parallel to the flow $\avg{\msymv_\|}$ [\fig\ref{fig:model}(b)] and bounded mean square displacement [\fig\ref{SIFig:MSD_Kurtosis}(a)~\cite{SI}]. This establishes the existence of a yield-stress-like response [\fig\ref{fig:model}(b)] and Movie S1~\cite{SI}], and we find deformability $d$ affects the threshold strongly, with $\text{Pe}^{*}\sim d^{-1}$ [inset of \fig\ref{fig:model}(b)].

\begin{figure*}[t]
	\centering
    \includegraphics[width=\textwidth]{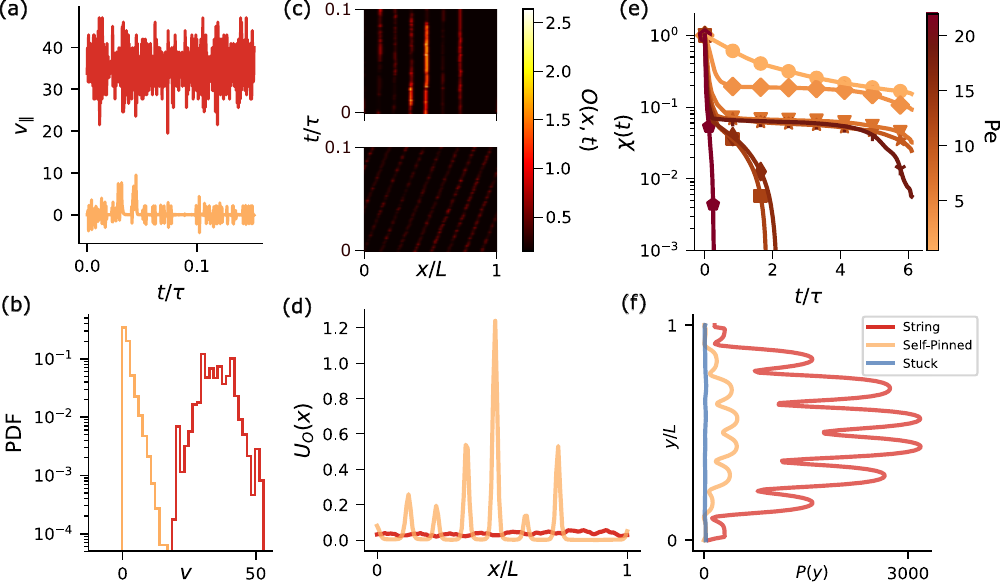}
	\caption{Stick-slip and pinning dynamics upon yielding. (a) Velocity time series $\msymv_\|(t)$ of a single droplet at two $\text{Pe}$ values: one showing stick-slip (orange; $\text{Pe} = 9.44$) and the other flowing dynamics (red; $\text{Pe} = 47.1$). $d = 0.1$ for both cases. (b) The probability density function (PDF) of droplet velocity $\msymv$ for these two cases, with excess kurtosis $\gamma_2 = 6.1$ (orange) and $0.6$ (red), and a larger $\gamma_2$ indicates further deviation from a normal distribution~\cite{SI}. (c),(d) Overlap profiles for a line of droplets in $x$ for the same $\text{Pe}$ values in (a),(b), as shown by (c) kymographs and (d) the time-averaged potential $U_O(x)$. (e) Time correlation of neighbours $\chi(t)$ for different $\text{Pe}$. (f) Momentum profile $P(y)$ for the three dynamical regimes at $d=0.5$ ($\text{Pe} = 0.62$ for stuck, $3.14$ for self-pinned, and $15.7$ for string-like). $L = 128$ for all panels but (e), where $L = 256$.}
    \label{fig2}
\end{figure*}

\begin{figure}[t]
    \centering
    \includegraphics[width=1.0\columnwidth]{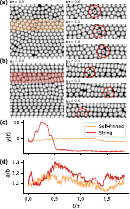}
    \caption{Structural behaviours differ between the self-pinned and string-like regimes. (a),(b) Snapshots tracking topological defects and their local packing (red star and circle) over time in (a) the stick-slip and (b) string-like regimes. In (a), defects advect with the suspension, whereas in (b) they are progressively annihilated through droplet deformations and rearrangements. (c),(d) Early-time dynamics showing (c) the transverse displacement and (d) the ratio of semi-major and minor axes of the defect droplets highlighted in (a),(b). Across all panels, $d = 0.5$ and $L = 256$,  and $\text{Pe} = 2.89$ for self-pinned and $\text{Pe} = 75.5$ for string-like.}
    \label{fig:fig3}
\end{figure}

\begin{figure}[t]
    \centering
    \includegraphics[width=\columnwidth]{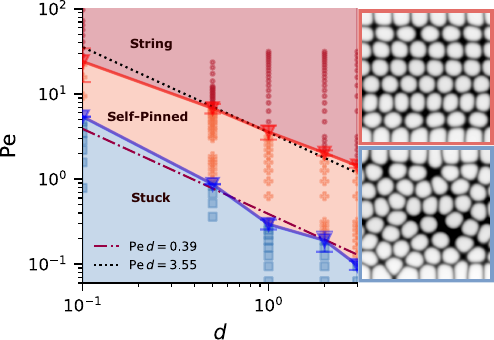}
    \caption{Phase diagram of the system across the parameter space ($d$, $\text{Pe}$). Left: Increasing $d$ and $\text{Pe}$ allows the system to transition from a stuck to a string-like phase through an intermediate self-pinned regime. The yielding transition (blue solid line) is defined based on the threshold $\avg{\msymv_\|} = 2.5\cdot10^{-3}$. The string-like transition (red solid line) is based on the threshold of the excess kurtosis of the velocity distribution, $\gamma_2 = 1$. Dashed lines represent the average $\text{Pe}\,d$ from points of each boundary line. Right: Representative snapshots of the system in the stuck and string-like phases.}
    \label{fig:phase_diagram}
\end{figure}

Importantly, yielding does not immediately lead to steady flow. Instead, just above $\text{Pe}^*$, droplet motion is highly intermittent: trajectories display long periods of arrest punctuated by sudden collective rearrangements [\fig\ref{fig2}(a) and Movie S2~\cite{SI}]. This stick-slip behaviour is reflected in the strongly non-Gaussian statistics of droplet velocities, as captured by the excess kurtosis $\gamma_2$, and the emergence of bimodality [\figs\ref{fig2}(b) and \ref{SIFig:MSD_Kurtosis}(b)~\cite{SI}], signalling the coexistence of slow and fast dynamical states. The mechanism underlying the stick-slip behaviour can be understood by analysing the dynamics of droplet overlaps. We focus on longitudinal overlaps $O(x,t)$ across a narrow slice of width $2R$ [see kymographs in \fig\ref{fig2}(c) and \cite{SI} for definition]. This overlap profile creates an effective rugged landscape, which we model via the time-average potential $U_O(x) = \avg{O(x,t)}_t$ [\figs\ref{fig2}(d) and \ref{SIFig:more_potentials}~\cite{SI}]. In this post-yielding regime, the system behaves as a ``self-pinned'' glassy material due to these persistent overlaps, which act as dynamically generated, effective quenched constraints. Because this landscape evolves on timescales much longer than individual droplet perturbations, it traps, or pins, the droplets. Motion then proceeds via activated-like events: droplets elastically load against these constraints before abruptly slipping to a new configuration, leading to the observed stick-slip dynamics (see Movie S2~\cite{SI}). 

Consistent with this picture, a key quantitative signature of the self-pinned glassy phase is the persistence of local environments, akin to the physics of caging in glasses~\cite{Sciortino2005,Zaccarelli2009}. By analysing neighbour correlation (see~\cite{SI}), we find that droplets remain caged by neighbouring overlaps over long timescales. Correspondingly, the typical timescale of neighbour exchange, quantified by the temporal correlation function $\chi(t)$, is much larger than the advective timescale \(\tau = L^2/(M\epsilon)\) [\fig\ref{fig2}(e)] and may display a plateau at long times, similar to the intermediate scattering function of colloidal glasses~\cite{Sciortino2005}.

Upon further increase in $\text{Pe}$, the suspension undergoes a second transition to a free-flowing, string-like state. In this regime, droplets deform sufficiently to continuously exchange neighbours, leading to a rapid decay of caging correlations [\fig\ref{fig2}(e) and Movie S3~\cite{SI}]. Accordingly, the effective pinning landscape is progressively erased: overlaps anneal over time, and motion becomes smooth and sustained [\fig\ref{fig2}(c)]. The sharp decrease in the caging correlation time and the qualitative difference in the overlap landscapes support the existence of a dynamical transition between the self-pinned glass and the flowing regime. 

The observations of three different regimes---solid-like, self-pinned glass, and flowing---can be captured qualitatively by an analogy with a modified Prandtl–Tomlinson model describing dry friction~\cite{Prandtl1928}. In our case, the role of the substrate potential is played not by an externally imposed landscape, but by the self-generated, slowly evolving potential arising from the network of overlaps, which we called $U_O$. In the solid phase, $U_O$ is constant in time and traps the droplets. In the self-pinned glass phase, $U_O$ evolves very slowly; it permits motion, but this requires overcoming local overlap barriers. This leads to long ``stick'' spells punctuated by sudden slips when the barriers drop enough for the forcing to drive the droplet through (\fig\ref{SIFig:more_potentials}~\cite{SI}). Finally, in the flowing phase, $U_O$ is approximately flat and no longer hinders motion. 

It is interesting to ask whether macroscopic and structural observables also differ near the glassy-to-flowing transition.  
Macroscopically, the self-pinned glass exhibits a different flow pattern compared with that in the flowing phase, as indicated by the averaged momentum of the suspension $P(y) = \frac{1}{LV}\int dx\sum_{i=1}^N\phi_i\msymv_{i,x}^{\text{CM}}$ [\fig\ref{fig2}(f)]. Rather than establishing a parabolic profile mirroring that of the advecting fluid, the suspension exhibits plug flow, where the bulk moves \textit{en masse} outside of thin boundary layers.

Concomitantly, the system undergoes a structural reorganisation, although this does not occur as a single sharp transition. Upon yielding, the self-pinned regime displays a partial increase in positional order, reflected in a finite value of the ``laning'' parameter $\Psi_L$ (\fig\ref{SIFig:laning}, see~\cite{SI} for definition). This parameter gradually increases as droplets form strings which flow smoothly. While $\Psi_L$ captures the build-up of alignment, it does not by itself sharply distinguish between the self-pinned glass and the fully flowing state. This is also the case for other structural observables, such as the bond-orientational order parameter $\Psi_6$ measuring hexatic order (\fig\ref{SIFig:structure}~\cite{SI}). 

The structural evolution is also evident in the behaviour of topological defects. In the self-pinned regime, five--seven disclination pairs in the local hexatic order are long-lived and move along the flow, reflecting the dynamically quenched nature of the underlying configuration [\fig\ref{fig:fig3}(a) and Movie S4~\cite{SI}]. In contrast, in the string-like flowing regime, these defects are progressively annihilated as droplets deform and reorganise, leading to a dynamically renewed structure [\fig\ref{fig:fig3}(b)]. Consistently, early-time transverse displacements along the flow-gradient direction [\fig\ref{fig:fig3}(c)] and the ratio between semi-major and minor axes of droplets [\fig\ref{fig:fig3}(d)] are significantly enhanced in the flowing regime, reflecting the increased mobility associated with neighbour exchange (Movie S5~\cite{SI}). The structural analysis here supports the scenario that while the yielding transition is associated with singularities in thermodynamic observables, the transition between the self-pinned glass and the string-like flowing phase is dynamical in origin. 

Based on the centre-of-mass velocity $\avg{\msymv_\|}$ and the excess kurtosis $\gamma_2$ of the velocity distribution, we build a phase diagram delineating the stuck, self-pinned glass, and string-like regimes within the parameter space ($d$,~$\text{Pe}$) (\fig\ref{fig:phase_diagram}; see \fig\ref{SIFig:finit_size_scaling} for finite-size scaling~\cite{SI}). This shows that the dynamical transition between glassy and flowing states also depends on $d$, and that the self-pinned phase exists for all $d$ probed. The shape of the phase diagram can be understood by noting that the physics is governed by the ratio between the timescales associated with overlap-driven motion and flow. As the former increases with $d$ and the latter decreases with $\text{Pe}$, a key dimensionless number is $\text{Pe}\,d$, giving a single parameter controlling the system state. Indeed, a numerical analysis of our phase diagram shows that, to a good approximation, if $\text{Pe} \,d \lesssim 0.39$ the system is stuck; if $0.39 \lesssim \text{Pe}\,d \lesssim 3.55$ it behaves as a self-pinned glass; whereas string-like flow arises when $\text{Pe}\,d \gtrsim 3.55$ (\fig\ref{fig:phase_diagram}). 

\textit{Discussions}---We have shown that a dry suspension of deformable droplets in pressure-driven flow undergoes two successive dynamical transitions controlled by forcing and deformability. At low forcing, the system behaves as an arrested amorphous solid. Above a first threshold, yielding occurs through intermittent stick-slip motion, leading to a dynamically arrested flowing state, which we interpret as a self-pinned glass. At larger forcing, the system undergoes a second transition to a continuously flowing, string-like regime characterised by rapid neighbour exchange and sustained transport.

Our results suggest that these two transitions arise from the interplay between droplet deformability and overlap-mediated interactions. Upon yielding, persistent overlaps evolve slowly and act as dynamically self-generated constraints, producing an effective rugged landscape that cages droplets over long timescales. Motion then proceeds through intermittent rearrangements reminiscent of the loading-and-release dynamics of the Prandtl--Tomlinson model near depinning. In contrast to standard depinning systems, however, the pinning landscape is not externally imposed but generated collectively by the suspension itself. Increasing forcing progressively anneals this overlap structure by enabling droplets to deform and exchange neighbours. This leads to a sharp reduction in caging correlation times and the disappearance of long-lived overlap patterns, allowing the system to transition from glassy, stick-slip motion to smooth, string-like flow. The weak structural signatures accompanying this transition, compared with the dramatic change in dynamical correlations, suggest that the glass-to-flowing transition is mainly dynamical in nature.

The mechanism identified here may extend beyond dry droplet suspensions. In particular, self-generated pinning and long-lived caging could arise generically in driven assemblies of deformable objects, including active matter~\cite{wiese2023,Sariyar2026} and dense cellular tissues~\cite{Loewe2020,henkes2020}. Hydrodynamic interactions, neglected in the present work, may further enrich this phenomenology by sustaining collective oscillations and long-range correlated rearrangements, providing an interesting direction for future work.

\textit{Acknowledgments}---We thank ESPRC for access to the HPC resources at EPCC (Cirrus). The authors would like to thank A. Morozov, T. Shendruk, and A. Bensabat for fruitful discussions.

\bibliography{refs}

\FloatBarrier
\clearpage
\pagebreak
\title{\papertitle \\ Supplemental Material}

\maketitle

\setcounter{equation}{0}
\renewcommand{\theequation}{S\arabic{equation}}

\setcounter{figure}{0}
\renewcommand{\thefigure}{S\arabic{figure}}

\setcounter{table}{0}
\renewcommand{\thetable}{S\arabic{table}}

\section{Phase Field Dynamics}
In our study, we consider a dry suspension of $N$ deformable particles, described as interacting phase fields. The particles interact via steric repulsion and are advected by a Poiseuille flow in the $x$ direction. The dynamics of the $i$th droplet is governed by the equation
\begin{equation}
    \frac{\pd\phi_i}{\pd t} + \vec{v}\cdot\vec{\nabla}\phi_i = M\nabla^2\frac{\delta\mathcal{F}}{\delta\phi_i} \;,
    \label{eqnSI:cahn}
\end{equation}
where $\mathcal{F}$ is the system's equilibrium free energy:
\begin{equation}
\label{eqn:free_energy_dry_droplet}
\begin{aligned}
\mathcal{F}
= \sum_{i=1}^N \Bigg[
& \int d^2\vec{r}
\left[
\frac{\alpha}{4}\phi_i^2(\phi_i-\phi_0)^2
+ \frac{\kappa}{2}(\vec{\nabla}\phi_i)^2
\right] \\
&+ \epsilon \sum_{i<j=1} \int d^2\vec{r} \, \phi_i^2 \phi_j^2
\Bigg] \;.
\end{aligned}
\end{equation}
We impose neutral wetting conditions, requiring that the following conditions are met for each droplet at the two walls ($y = 0$ and $y = L$) confining our system:
\begin{equation}
    \begin{aligned}
        \left.\frac{\pd\mu_i}{\pd x}\right|_{y=0,L} = 0 \;,\\
        \left.\frac{\pd\nabla^2\phi_i}{\pd y}\right|_{y=0,L} = 0 \;,
    \end{aligned}
\end{equation}
where the first line ensures density conservation and the second sets the wetting to be neutral. We employ periodic boundary conditions in $x$ along the flow direction. 

\section{Model Implementations}
Here we describe the details of our numerical solution of \eqn\eqref{eqnSI:cahn}.
Evaluating the functional derivative of $\mathcal{F}$, \eqn\eqref{eqnSI:cahn} can be written as
\begin{equation}
    \begin{aligned}
       \frac{\pd\phi_i}{\pd t} = -\vec{v}\cdot\vec{\nabla}\phi_i + M \Bigg[ \alpha \left( \phi^3_i - \frac{3}{2}\phi_i^2 \phi_0 + \frac{1}{2} \phi_i \phi_0^2\right) \\
       -\kappa \nabla^2 \phi_i + 2\epsilon \phi_i\left[h(\vec{r})-\phi_i^2\right] \Bigg] \;,
    \end{aligned}
    \label{eqn:decoupled}
\end{equation}
where we have introduced an auxiliary field $h(\vec{r}) = \sum_{j=1}^N \phi_j^2(\vec{r})$ that enables one to decouple and parallelise the computation of individual phase fields \cite{Nonomura2012}. Specifically, we first calculate $h(\vec{r})$ using the phase fields at the current timestep, and then perform the update of individual phase fields in parallel with $h(\vec{r})$ known.

We simulate \eqn\eqref{eqn:decoupled} using a finite-difference method. Length is expressed in terms of the lattice spacing $\Delta x$, and time in units $\Delta t \equiv \lambda \delta t$, where $\delta t$ is the size of each timestep and $\lambda$ a scale factor (we use different $\delta t$ for different deformabilities $d$ to keep the numerical integration stable; see Table~\ref{tab:sim_params}). The simulation code is written in a mixture of \texttt{C} and \texttt{C++} and is parallelised using \texttt{OpenMP}. In line with previous studies~\cite{Mueller2019,Nonomura2012}, we employ subdomain decomposition to reduce computational costs, whereby we only resolve each field within a square subdomain with linear dimension $L_{\text{sub}} \leq L$. Here we set $L_{\text{sub}} = 128$, which is the smallest system size investigated and is much larger than the droplet radius $R$. The subdomain acts as a co-moving frame---i.e., we keep each droplet at the centre of its subdomain by performing a shifting algorithm that moves the droplet back to the subdomain centre when it has migrated more than two lattice units in either direction. This allows us to solve \eqn\eqref{eqn:decoupled} using fixed boundary conditions ($\phi_i|_{\pd\Omega_i} = 0$) in these subdomains (while respecting the global boundary conditions at the walls $y = 0$ and $y = L$).

We initialise the system by randomly placing $N$ circular droplets of radius $R = 8$ in the simulation box (with $\phi = 2$ within the droplet). We then allow them to relax for $10^4\Delta t$ (with $\text{Pe} = 0$) to remove any large overlaps. Next, we switch on the flow profile $v_x(y)$ and evolve the droplets for time $t_{\text{sim}} = 10^7\Delta t$, with positions and overlaps sampled every $10^3\Delta t$. The full set of parameter values used in the simulations, including the simulation box size $L$ and packing fraction $\Phi$, is summarised in Table~\ref{tab:sim_params}. For each set of parameters we averaged between $5$ \& $10$ simulations.

\begin{table*}[t]
\begin{ruledtabular}
\begin{tabular}{cccc}
\textrm{Parameter}&
\textrm{Interpretation}&
\textrm{Dimensions}&
\textrm{Value(s)}\\
\colrule
    $d$ & Deformability & - & $0.05$, $0.1$, $0.5$, $1$, $2$, $3$ \\
    $R$ & Droplet radius & [L] & $8$ \\
    $\xi$ & Droplet interface thickness & [L] & $2$ \\
    $\epsilon$ & Droplet-droplet repulsion & [E][L]$^{-2}$ & $0.05$ \\
    $M$ & Mobility & [L]$^4$[E]$^{-1}$[T]$^{-1}$ & $0.1$ \\
    $v_{\text{max}}$ & Maximum flow speed & [L][T]$^{-1}$ & $0$--$6\cdot 10^{-6}$ \\
    $\lambda$ & Timescale factor & - & $8~(d = 0.05)$, $4~(d = 0.1)$, $2~(d = 0.5)$, $1~(d=1,2,3)$ \\
    $\delta t$ & Timestep & [T] &$0.125~(d = 0.05)$, $0.25~(d = 0.1)$, $0.5~(d = 0.5)$, $1~(d=1,2,3)$ \\
    $\Delta t$ & Time unit & [T] & $1$ \\
    $\Delta x$ & Lattice size & [L] & $1$ \\
    $L_{\text{sub}}$ & Subdomain size & [L] & $128$ \\
    $L$ & Simulation box size & [L] & $128$, $192$, $256$, $512$ \\
    $\Phi$ & Packing fraction & - & $0.67$ \\
\end{tabular}
\end{ruledtabular}
\caption{\label{tab:sim_params}Parameter values used in the simulations.}
\end{table*}

\section{Dynamical Observables}
We consider several dynamical observables to quantify the change in the system behaviour as we increase the forcing $\text{Pe}$ driving the flow and droplet deformability $d$. The following discusses the definition of each observable. 

\subsection{Centre-of-Mass Velocity}
A direct measure of the system dynamics is the centre-of-mass velocity of each droplet. In components, say for the $i$th droplet, they are given by
\begin{equation}
    {\msymv}_{i,x}^{\text{CM}}(t) = \frac{x_i(t+\Delta t)-x_i(t)}{\Delta t} \;,
\end{equation}
\begin{equation}
    {\msymv}_{i,y}^{\text{CM}}(t) = \frac{y_i(t+\Delta t)-y_i(t)}{\Delta t} \;.
\end{equation}
The mean centre-of-mass velocity parallel to the flow across all droplets is therefore
\begin{equation}
    {\msymv}_\|(t) = \frac{1}{NV}\sum_{i=1}^N\msymv_{i,x}^{\text{CM}}(t) \;,
\end{equation}
while the mean velocity magnitude is
\begin{equation}
    {\msymv}(t) = \frac{1}{NV}\sum_{i=1}^N\sqrt{\msymv_{i,x}^{\text{CM}}(t)^2+\msymv_{i,y}^{\text{CM}}(t)^2} \;.
\end{equation}
We have normalised these quantities over $V = M\epsilon/L$, a typical velocity of the droplets arising from passive thermodynamic forces. We report $\avg{\msymv_\|}$ when exploring the dynamics at different parameter points $(d,\text{Pe})$, where $\avg{\dots}$ is an ensemble average over time and simulation runs. In particular, the average over time is taken over all the steady-state timesteps (i.e., $t \geq t_{\text{eq}} \equiv  6\cdot10^6 \Delta t$). We consider the system yielded once $\avg{\msymv_\|} \geq 2.5\cdot10^{-3}$ and remains above this threshold with increasing $\text{Pe}$. 

\subsection{Mean Square Displacement}
We also compute the mean square displacement (MSD) of the droplets as
\begin{equation}
    \text{MSD}(t) = \frac{1}{N}\avg{\sum_{i=1}^N\, [\vec{r}_i(t+{\tau}) - \vec{r}_i(\tau)]^2}_\tau \;,
\end{equation}
where $\vec{r}_i$ is the centre of mass of the $i$th droplet in the lab frame. Figure \ref{SIFig:MSD_Kurtosis}(a) shows the long-time MSD [i.e., $\text{MSD} = \text{MSD}(t_{\text{sim}})$] for different deformabilities $d$ across a range of $\text{Pe}$ values. Below the yielding transition, the MSD remains unchanged with increasing $\text{Pe}$; it then sharply increases once $\text{Pe}$ exceeds a critical value $\text{Pe}^*$. Similar to Figure~\ref{fig:model}, increasing $d$ facilitates yielding, with lower $\text{Pe}^*$ required.

\subsection{Excess Kurtosis of the Velocity Distribution}
\begin{figure}[t]
    {\includegraphics[width=\columnwidth]{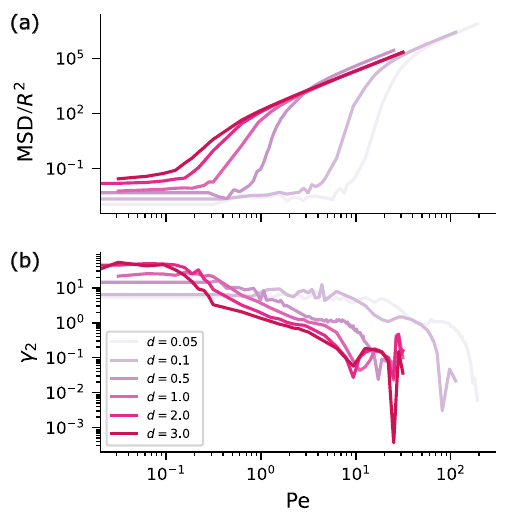}}
	
    \caption{Dynamical observables capturing the phase transitions. (a) Mean square displacement ($\text{MSD}$) of droplets as a function of Péclet number ($\text{Pe}$) for deformabilities $d = 0.05$ to $3.0$. (b) Excess kurtosis $\gamma_2$ of the droplet velocity $\msymv$ distribution as a function of $\text{Pe}$ for the same range of $d$. The system size is $L = 128$ for both panels.}
    \label{SIFig:MSD_Kurtosis}
\end{figure}

The centre-of-mass velocity and MSD of the suspension allow us to pinpoint the yielding transition; however, they provide limited information about the dynamical behaviour of the system beyond this transition, especially in quantifying the observed stick-slip dynamics and the transition to string-like motion. To this end, we examine the time series of the centre-of-mass velocity $\msymv$ and its distribution, which shows non-Gaussian-like behaviour in the stick-slip regime [see \fig\ref{fig2}(b)]. This motivates us to measure the excess kurtosis $\gamma_2$ of the distribution, which is defined as
\begin{equation} 
\gamma_2 = \frac{\mu_4}{\sigma^4} - 3 \;,
\end{equation}
where $\mu_4$ and $\sigma$ are the fourth standard moment and the standard deviation, respectively, of the distribution. This metric captures the degree of tailedness of a distribution relative to that of a normal distribution, which has $\gamma_2 = 0$. Hence, when $\gamma_2 > 0$, the distribution has fatter tails than that of a Gaussian, providing a clear metric for non-Gaussian-like behaviour, such as bimodality. Figure~\ref{SIFig:MSD_Kurtosis}(b) shows $\gamma_2$ versus $\text{Pe}$ for different deformabilities. $\gamma_2$ captures the yielding transition as we see an initial decrease once $\text{Pe} > \text{Pe}^*$. Setting a threshold at $\gamma_2 = 1$ allows us to capture the dynamical transition from stick-slip to string-like motion and therefore build the phase diagram in Figure~\ref{fig:phase_diagram}.

\begin{figure}[t]
    \centering
    \includegraphics[width=\columnwidth]{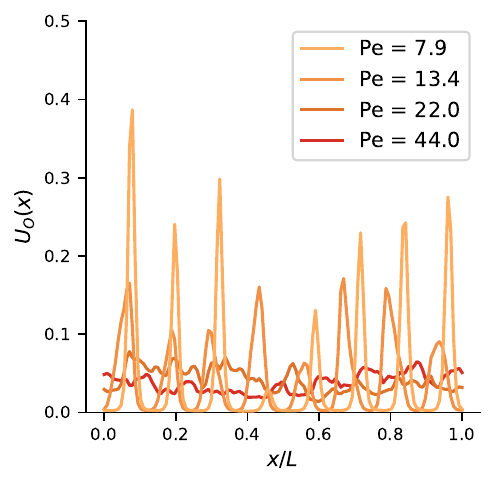}
    \caption{Profiles of the overlap-induced potential $U_O$ as a function of $x$ for different $\text{Pe}$ values. The system size is $L = 128$ and $d = 0.1$.}
    \label{SIFig:more_potentials}
\end{figure}

\subsection{Overlap-Induced Potential}
In the main text, we suggest that the stick-slip motion arises from a ``self-pinning'' phenomenon, in which the persistence of droplet-droplet overlaps generates an effective rugged profile, driving glassy dynamics. We measure this profile by calculating longitudinal overlaps
\begin{equation}
    O(x,t;y_0) = \frac{\epsilon}{2R}\sum_{i=1}^N\sum_{i<j=1}^N\int_{y_0}^{y_0+2R}dy\,\phi_i^2\phi_j^2 \;,
\end{equation}
where we perform the integral over a vertical slice of the box corresponding to the droplet diameter $2R$. $y_0$ is chosen to coincide with the bottom of one of the lanes, and changing $y_0$ does not alter the trend of how the profile varies with $\text{Pe}$. From the overlap profile, we further define the time-average potential $U_O(x) = \avg{O(x,t)}_t$. In Figure~\ref{SIFig:more_potentials}, we show further examples of this potential for different $\text{Pe}$ values. At low $\text{Pe}$, the profiles are rugged with pronounced peaks, reflecting spatially localised overlaps consistent with the self-pinned regime. Increasing $\text{Pe}$ progressively smooths $U_O(x)$, indicating reduced overlap localisation and more uniform transport.

\subsection{Velocity Alignment Order Parameter}
\begin{figure}[t]
    \centering
    \includegraphics[width=\linewidth]{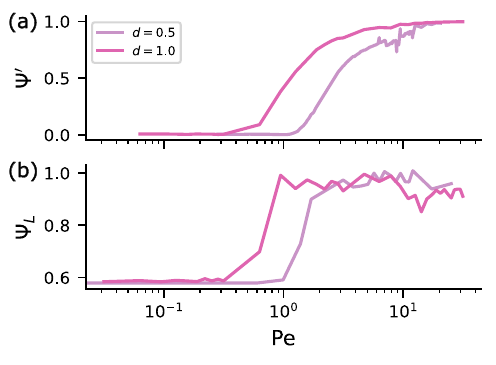}
    \caption{Other dynamical and structural observables quantifying the phase transitions. (a) Velocity alignment order parameter $\Psi'$ and (b) laning parameter $\Psi_L$ as a function of $\text{Pe}$ for $d = 0.5$ and $d = 1.0$. The system size is $L = 128$ for both panels.}
    \label{SIFig:laning}
\end{figure}

The aforementioned observables allow us to capture the three dynamical regimes stated in the phase diagram. However, these observables reveal little about the internal dynamics---i.e., how droplets move within the suspension relative to those around them. The local surroundings of a droplet strongly affect its behaviour; for instance, topological defects lead to larger unfavourable local overlaps~\cite{Loewe2020}, while coordinated movement of droplets can drive defect annihilation. Therefore, it is important to understand how the droplets move relative to each other. To quantify this, we introduce the velocity alignment order parameter
\begin{equation}
    \Psi' = \abs{\avg{\frac{1}{N}\sum_{j=1}^N \frac{1}{N_{\text{nn},j}}\sum_{k \in \text{nn}} \exp\left(i\varphi_{jk}\right)}} \;,
\end{equation}
where $\varphi_{jk}$ is the angle between the centre-of-mass velocity vectors of droplets $j$ and $k$. Here, the sum over $k$ runs over the nearest neighbours (nn) of $j$, and $N_{\text{nn},j}$ gives its total number of neighbours. Unless specified otherwise, we use Delaunay triangulation to determine nearest neighbours (this also applies to other observables that require nearest-neighbour identification). The value of $\Psi'$ ranges from $0$ to $1$, with $\Psi' = 1$ indicating that the droplets' velocity vectors are locally aligned. Figure~\ref{SIFig:laning}(a) shows how $\Psi'$ varies with $\text{Pe}$ for two different deformabilities. As $\text{Pe}$ increases, we see that the velocity vectors become increasingly aligned at a local level, consistent with the picture that the suspension flows more smoothly with larger forcing.

\section{Structural Observables}
While the transitions discussed in this work are dynamical in nature, it is also interesting to examine the structural changes as the system transitions between different regimes. In the following, we discuss several structural observables that we have analysed.

\subsection{Laning Parameter}
A notable feature observed upon yielding is that the droplets rearrange themselves into lanes. These lanes become progressively more well-defined as the system moves deeper into the string-like, flowing regime. We note that the lanes formed are not evenly spaced in $y$ due to the differential shear stress and that the droplets are deformable, resulting in narrower spacing between lanes near the wall than near mid-channel [see \fig\ref{fig:fig3}(b)]. To construct a parameter measuring the degree of aperiodic ``laning'', we observe that when lanes are perfect, each lane will have $n = 2RN/L$ droplets. By comparing the actual number of droplets sharing the same lane with this ideal number $n$, we define the laning parameter as
\begin{equation}
    \Psi_{L} = \avg{\frac{1}{N}\sum_{i=1}^N\sum_{j\neq i}\frac{\Theta(\delta-\abs{y_{ij}})}{n-1}} \;,
\end{equation}
where $y_{ij}$ is the vertical separation between droplets $i$ and $j$, $\delta = \Delta x$ is the threshold for considering droplets to be in the same lane, and $\Theta$ is the Heaviside step function, with $\Theta(x) = 1$ if $x > 0$ and $0$ otherwise. The $-1$ term in the denominator excludes self-comparison. We note that $\Psi_L$ ranges roughly from 0 to 1, where a value near 1 indicates that all droplets are in their respective lane. 

Figure~\ref{SIFig:laning}(b) displays $\Psi_L$ as a function of $\text{Pe}$ for two different deformabilities. Similar to $\Psi'$, it remains near zero when the system is stuck and increases steeply as the system yields and becomes self-pinned, before finally reaching a plateau in the string-like regime. Both $\Psi_L$ and $\Psi'$ indicate that the yielding transition shifts to lower $\text{Pe}$ with increasing $d$, consistent with the phase diagram (\fig\ref{fig:phase_diagram}). Compared to $\Psi'$, the increase of $\Psi_L$ near yielding is sharper, highlighting the structural nature of the yielding transition.

\begin{figure}[t]
    \centering
    \includegraphics[width=\linewidth]{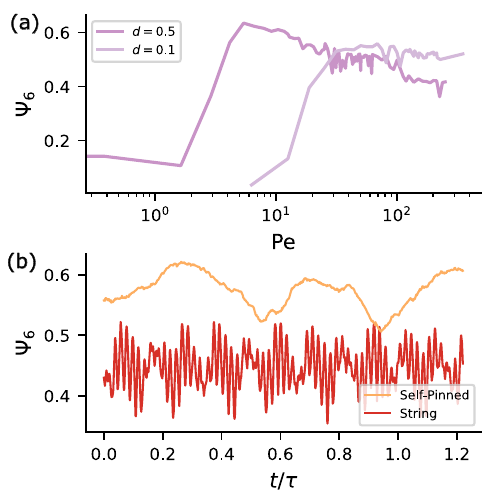}
    \caption{Examining the structure of the suspension with the bond-orientational order parameter $\Psi_6$. (a) $\Psi_6$ as a function of $\text{Pe}$ for $d = 0.1$ and $d = 0.5$. (b) Time series of $\Psi_6$ for the self-pinned glass ($\text{Pe} = 4.2$) and string-like flowing phases ($\text{Pe} = 125.0$) with $d = 0.5$. The system size is $L = 256$ for both panels.}
    \label{SIFig:structure}
\end{figure}

\subsection{Bond-Orientational Order Parameter}
Apart from laning, we examine the degree of local hexatic order by measuring the bond-orientational order parameter. The local order for each droplet $j$ is given by
\begin{equation}
    \psi_{6,j}(t) = \frac{1}{N_{\text{nn},j}}\sum_{k\in\text{nn}} \text{exp}\left(6i\theta_{jk}\right) \;,
\end{equation}
where $\theta_{jk}$ is the angle between the $x$-axis and the vector connecting the centres of mass of droplets $j$ and $k$. We compute both time series of the global order
\begin{equation}
    \Psi_6(t) = \abs{\frac{1}{N}\sum_{j=1}^N\psi_{6,j}(t)}
\end{equation}
and the ensemble-averaged order parameter
\begin{equation}
    \Psi_6 = \abs{\avg{\frac{1}{N}\sum_{j=1}^N\psi_{6,j}(t)}} \;.
\end{equation}
Here, $\Psi_6 \simeq 1$ corresponds to droplets having nearest neighbours close to a perfect hexagonal arrangement, whereas $\Psi_6 \simeq 0$ when this orientational order is lost.

Figure~\ref{SIFig:structure} shows $\Psi_6$ as we vary $\text{Pe}$ and the time series $\Psi_6(t)$ for an example in the self-pinned phase and another in the string-like phase. We find $\Psi_6$ sharply increases upon yielding (i.e., $\text{Pe} > \text{Pe}^*$), before decaying progressively as the suspension transitions from the self-pinned glass to the string-like flowing phase. This decrease is partly due to the loss of local, regular six-fold packing as droplets form lanes with aperiodic separation. Additionally, we note that in the string-like phase $\Psi_6$ exhibits pronounced oscillations, reflecting the periodic reorganisation of local structure as droplets continuously exchange neighbours; however, in the self-pinned glass phase it remains comparatively steady, consistent with the dynamically quenched nature of that regime.

\subsection{Nearest-Neighbour Correlation}
Another aspect that allows us to delineate the self-pinned and string-like regimes is the dynamics of neighbour exchanges, especially the correlation of nearest neighbours over time. To measure this, we construct a matrix that indicates whether a droplet (say $j$) is a nearest neighbour of another (say $i$) at time $t$ as
\begin{equation}
    P_{ij}(t) = 
    \begin{cases}
        \Theta(\abs{y_{ij}}-\delta_{\text{cross}}) & \text{if } j \in \text{nn}_i \;, \\
        0 & \text{otherwise} \;,
    \end{cases}
\end{equation}
where $\delta_{\text{cross}} = R/2$ is the threshold for distinguishing in-lane and out-of-lane neighbours (we exclude in-lane neighbours as they are unlikely to change over time). This matrix allows us to define a time correlation function of neighbour exchanges as
\begin{equation}
    \chi(t) = \avg{\frac{1}{N}\sum_{j=1}^N P_{ij}(t) P_{ij}(t-\Delta t) \dots P_{ij}(t_0)} \;,
\end{equation}
where the average $\avg{\dots}$ is taken over droplets and start time $t_0$. In essence, $\chi(t)$ measures how likely two droplets remain in contact at future times if they are out-of-lane neighbours at time $t_0$. Measurements of this correlation function are reported in Figure~\ref{fig2}(e).

\begin{figure}[t]
    {\includegraphics[width=1\columnwidth ]{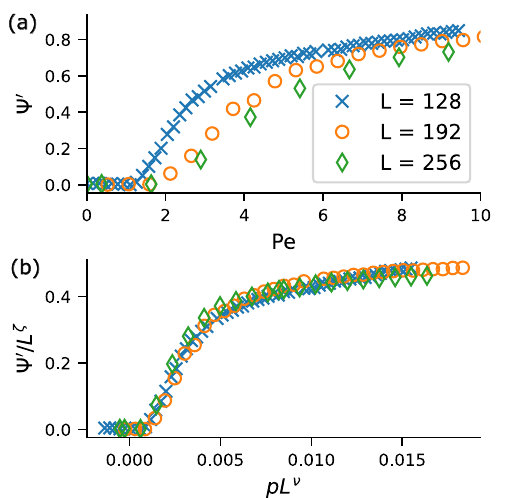}}
	\caption{Finite-size scaling analysis of the yielding transition. (a) Velocity alignment order parameter $\Psi'$ as a function of $\text{Pe}$ for system sizes $L = 128$, $192$, and $256$ with $d = 0.5$. (b) Collapse of the data for $\Psi'$ from different system sizes $L$ based on our estimate of the scaling parameters ($\text{Pe}^*$, $\nu$, $\zeta$), which are obtained from minimising the residual $\mathcal{R}$.}
    \label{SIFig:finit_size_scaling}
\end{figure}

\section{Finite-Size Scaling}
The existence of a stick-slip, intermittent state after yielding prompts us to further investigate the nature of this transition. To this end, we perform finite-size scaling at $d=0.5$ for systems of sizes $ L=128$, $192$, and $256$. Using $\Psi'$ as the order parameter for yielding, we consider the scaling ansatz
\begin{equation}
    \Psi' = L^\zeta f(pL^\nu) \;,
\end{equation}
where $\zeta$ and $\nu$ are scaling exponents to be determined and $p = \text{Pe}/\text{Pe}^* -1$, with $\text{Pe}^*$ the critical P{\'e}clet value as $L \xrightarrow{}\infty$.
We estimate $\text{Pe}^*$, $\nu$, and $\zeta$ numerically by means of the procedure outlined in \cite{Somendra2001}.
Specifically, we compute a residual function $\mathcal{R}$ that measures the pairwise differences between data sets from different $L$ when scaled by a given set of $(\text{Pe}^*, \nu, \zeta)$:
\begin{equation}
    \mathcal{R} (\text{Pe}^*, \nu,\zeta) = \frac{1}{\mathcal{N}} \sum_i \sum_{j \neq i} \sum_{k, \text{over}} \left | \Psi'_{jk}L_j^{-\zeta} - \mathcal{L}_i (p_{jk}L^\nu_j)\right| \;,
\end{equation}
where the first two sums are over all possible pairs of data sets $i$, $j$, and the innermost sum is carried out over the data points $k$ of set $j$ that are within the rescaled domain of set $i$. $\mathcal{L}_i(x)$ is an interpolating function based on the values of set $i$, and $\mathcal{N}$ is the number of data points compared in the sums. Minimising $\mathcal{R}$ with the Nelder-Mead algorithm~\cite{Loewe2020}, we find the parameters that best collapse our data sets are as follows:
\begin{center}
     $\text{Pe}^* = 0.763(9)$\\
     $\nu = -1.36(1)$\\
     $\zeta = 0.215(7)$ \;. 
\end{center}

\section{Phase Diagram Construction}
We discuss here in more detail the construction of the phase diagram shown in Figure~\ref{fig:phase_diagram}. As mentioned in the main text, the yielding transition is defined based on the threshold $\avg{\msymv_\|} = 2.5\cdot10^{-3}$, while the boundary separating the self-pinned and string-like phases is based on the excess kurtosis of the velocity distribution $\gamma_2 = 1$. More specifically, to construct each boundary, we examine the observable value (i.e., $\avg{\msymv_\|}$ or $\gamma_2$) across all $\text{Pe}$ for each $d$ and find the two points nearest to the threshold. We then use the linear interpolation of these two points to evaluate the precise $\text{Pe}^*$ when the threshold is crossed. To estimate the errors of the drawn boundary, we perform bootstrapping on the observable data and repeat the boundary estimation $500$ times. The line drawn represents the mean of the sampled boundaries, with errors denoting the standard deviation of the sampling.

In the main text, we suggest that the transition between phases is related to the competition of the timescales associated with overlap-induced motion and advection, giving rise to a single control parameter $\text{Pe}\,d$, with the estimated boundary lines supporting this conjecture (see \fig\ref{fig:phase_diagram}). Here, we conduct a dimensional analysis to show that the ratio of these timescales is proportional to $\text{Pe}\,d$. First, we observe that the timescale of overlap-induced motion is related to the time for a deformed droplet to relax and return to an undeformed, circular shape. Clearly, this relaxation is proportional to the inverse mobility $M^{-1}$ (with dimensions [E][T][L]$^{-4}$) and should be faster for droplets with higher interfacial tension $\sigma$ (which has dimensions [E][L]$^{-1}$); hence, we hypothesise the relaxation time $\tau_{\text{relax}}$ to take the form
\begin{equation}
    \tau_{\text{relax}} = \frac{\ell_1\ell_2\ell_3}{M\sigma} \;,
\end{equation}
where $\ell_i$ ($i = 1,2,3$) represent some unknown lengthscales (which, for instance, could be related to the droplet radius $R$). Next, the advection time is simply given by
\begin{equation}
    \tau_{\text{flow}} = \frac{L}{v_{\text{max}}} = \frac{L^2}{\text{Pe}\,M\epsilon} \;,
\end{equation}
where we have applied our definition of the P{\'e}clet number. Taking the ratio of the two timescales and recalling that the interfacial tension is $\sigma = \sqrt{8\kappa\alpha/9}$, we find
\begin{equation}
\begin{split}
    \frac{\tau_{\text{relax}}}{\tau_{\text{flow}}} &= 
    \frac{\ell_1\ell_2\ell_3}{L^2} \frac{\text{Pe}\,\epsilon}{\sigma} \\
    &=\frac{3\ell_1\ell_2\ell_3}{2L^2}\left(\frac{\alpha}{2\kappa}\right)^{1/2}\,\text{Pe}\,\frac{\epsilon}{\alpha} \\
    &= \frac{3\ell_1\ell_2\ell_3}{2L^2\xi}\,\text{Pe}\,d \\
    &\propto \text{Pe}\,d \;,
\end{split}  
\end{equation}
where in the third equality we have used the definition of deformability and the fact that the interfacial thickness of a droplet is $\xi = \sqrt{2\kappa/\alpha}$. This thus demonstrates that $\text{Pe}\,d$ captures the competition of the two timescales up to a dimensionless prefactor.

\section{Simulation Movies}
We provide below the captions of the simulation movies. For all movies, the system size is $L = 256$.
\begin{itemize}
    \item\textbf{Movies S1--S3}: Example simulation runs showing the dynamics of the system in all three phases: stuck (S1), self-pinned (S2), and string-like (S3). The movies begin at the point when the flow has just been switched on. All movies are for a system with $d = 0.5$ and $\text{Pe} = 0.37$, $1.89$, $62.9$, respectively. A red star highlights a reference droplet to guide the eye.
    
    \item\textbf{Movie S4}: A simulation run showing the presence of long-lived topological defects in the self-pinned glass phase. The parameters are $d = 0.5$ and $\text{Pe} = 2.89$ with a droplet trapped in a topological defect, highlighted with a red star.

    \item\textbf{Movie S5}: A simulation run showing the formation of lanes in the string-like regime. The strong shear flow facilitates droplet deformation, enabling long-lived topological defects to be annihilated (see the droplet highlighted with a red star). The parameters are $d = 0.5$ and $\text{Pe} = 75.5$. 
\end{itemize}

\end{document}